\documentclass{article}
\usepackage[cp1251]{inputenc}
\usepackage[english]{babel}
\usepackage{amssymb}
\def\<{\langle}
\def\>{\rangle}

\newcommand{\Lac}{{\Lambda}_c }

\newcommand{\ksi}{{\xi}_i}

\newcommand{\tks}{{\tilde \xi}}

\newcommand{\nui}{{\nu}_i}

\newcommand{\F}{{\Phi}}

\def\thod{\theta^{12}}
\def\thdo{\theta^{21}}

\newcommand{\eti}{{\eta}_i}

\newcommand{\etiz}{{\eta}_i^0}
\newcommand{\etit}{{\eta}_i^3}

\newcommand{\pstf}{\psi_{\tilde f} }

\newcommand{\Ps}{\Psi}
\newcommand{\Psz}{\Psi_0}
\newcommand{\Pszo}{\Psi_0^1}
\newcommand{\Pszi}{\Psi_0^i}

\newcommand{\vf}{\varphi }
\newcommand{\vfx}{\varphi\,(x) }

\newcommand{\vfytsp}{\varphi\,(y_1)\,\ts \cdots \ts \, \varphi\,(y_m) }

\newcommand{\vff}{\varphi_f }

\newcommand{\vffp}{\varphi_{f_1}\,\ts  \cdots \ts \,\varphi_{f_n}}
\newcommand{\vffkp}{\varphi_{f_k}\,\ts  \cdots \ts \,\varphi_{f_1}}
\newcommand{\vffnp}{\varphi_{f_n}\,\ts  \cdots \ts \,\varphi_{f_1}}
\newcommand{\vffkpop}{\varphi_{f_{k + 1}}\,\ts  \cdots \ts \,\varphi_{f_n}}

\newcommand{\vftf}{\varphi_{\tilde f} }
\newcommand{\vftfo}{\varphi_{\tilde f_1}}

\newcommand{\vftfm}{\varphi_{\tilde f_m}}

\newcommand{\xz}{x^0}
\newcommand{\xo}{x_1}
\newcommand{\xd}{x_2}
\newcommand{\xt}{x^3}
\newcommand{\xch}{x_4}
\newcommand{\xit}{x_i}
\newcommand{\xiz}{x_i^0}

\newcommand{\xitr}{x_i^3}

\newcommand{\xitz}{x_i^0}
\newcommand{\xito}{x_i^1}
\newcommand{\xitd}{x_i^2}
\newcommand{\xitt}{x_i^3}
\newcommand{\xj}{x_j}
\newcommand{\xjz}{x_j^0}

\newcommand{\xjt}{x_j^3}

\newcommand{\wf} {W\,(x_1,  \ldots, x_n)}

\newcommand{\wfo}{W\,(x_n,  \ldots,  x_1)}
\newcommand{\wfxi}{W\,(x_1,   \ldots, x_i,  x_{i +1}, \ldots,  x_n)}
\newcommand{\wfxio}{W\,(x_1,   \ldots, x_{i + 1}, x_i, \ldots,  x_n)}

\newcommand{\wfnx}{W\,(\nu_1,   \ldots, \nu_{n - 1}, X)}

\newcommand{\wfksx}{W\,(\xi_1,   \ldots, \xi_{n - 1}, X)}
\newcommand{\vfxtsp}{\varphi \,(x_1)\,\tilde \star\, \cdots  \tilde \star\, \varphi\,(x_n) }

\newcommand{\vfitvx}{{\varphi}_i \,(t, \vec{x}) }

\newcommand{\wfts}{W_{\tilde \star}\,(x_1, x_2,  \ldots, x_n)}

\newcommand{\Psl}{\langle \Psi}

\newcommand{\Pszl}{\langle \Psi_0}
\newcommand{\Pszr}{\Psi_0 \rangle}

\newcommand{\Phr}{\Phi \rangle}
\newcommand{\ts}{\tilde \star }

\newcommand{\dxo}{d\,x_1}

\newcommand{\dxch}{d\,x_4}
\newcommand{\dxn}{d\,x_n}

\newcommand{\cpn}{P_n}
\newcommand{\cpnz}{P_n^0}
\newcommand{\cpnt}{P_n^3}
\newcommand{\vcpn}{\vec{P_n}}
\newcommand{\ctn}{T_n}
\newcommand{\ctnmi}{T_n^-}
\newcommand{\bcvpl}{\bar{V^+}}
\newcommand{\cvdpl}{{V_2^+}}
\newcommand{\bcvdpl}{\bar{V_2^+}}

\newcommand{\fsi}{f_i}
\newcommand{\fsipo}{f_{i+1}}
\newcommand{\fsx}{f\,(x)}

\newcommand{\fsbxp}{f_1\,(x_1)\,\ts  \cdots   \ts \,{f_n\,(x_n)}}

\newcommand{\complessi}{\hbox{\rm I\hskip-5.9pt\bf C}}

\newcommand{\identity}{\hbox{\rm I\hskip-5.9pt\bf I}}
\begin{document}

\begin{center}
{ \Large{\bf{Classical Theorems in Noncommutative\\ Quantum Field
Theory }}}
\end{center}

\vspace{5mm}

\begin{center}
{\large \bf M. Chaichian$^a$,  M. Mnatsakanova$^b$, A. Tureanu$^a$
 and Yu.~ Vernov$^c$}

{\it $^a$High Energy Physics Division, Department of Physical
Sciences, University of Helsinki and Helsinki Institute
of Physics, P.O. Box 64, 00014 Helsinki, Finland\\
$^b$Skobeltsyn Institute of Nuclear Physics, Moscow State
University, Moscow, Russia \\
$^c$Institute for Nuclear Research, RAS, Moscow, Russia }
\vspace{5mm}


\end{center}

\vspace{5mm}
\begin{abstract}
 {Classical results of the axiomatic quantum field theory - Reeh and Schlieder's theorems,
irreducibility of the set of field operators and generalized Haag's
theorem are proven in $S \, O \, (1,1)$  invariant quantum field
theory, of which an important example is noncommutative  quantum
field theory. In $S \, O \, (1,3)$  invariant theory new
consequences of generalized Haag's theorem are obtained. It has been
proven that the equality of four-point Wightman functions in two
theories leads to the equality of elastic scattering amplitudes and
thus the total cross-sections in these theories.}
\end{abstract}

\section{Introduction}
Quantum field theory (QFT) as a mathematically rigorous and
consistent theory was formulated in the framework of the axiomatic
approach in the works of Wightman, Jost, Bogoliubov, Haag and others
(\cite{SW} - \cite{Haag}).

Within the framework of this theory on the  basis of  most general
principles such as Poincar\'{e} invariance, local commutativity and
spectrality, a number of fundamental physical results, for example,
the CPT-theorem and the spin-statistics theorem  were proven
\cite{SW} - \cite{BLT}.

Noncommutative quantum field theory (NC QFT) being one of the
generalizations of standard QFT has been intensively developed
during the past years (for reviews, see \cite{DN, Sz}). The idea of
such a generalization of QFT ascends to Heisenberg and it was
initially developed in Snyder's work \cite{Snyder}. The present
development in this direction is connected with the construction of
noncommutative geometry \cite{Connes} and new physical arguments in
favour of such a generalization of QFT \cite{DFR}. Essential
interest in NC QFT is also due the fact that in some cases it is a
low-energy limit of string theory \cite{SeWi}. The simplest and at
the same time most studied version of noncommutative field theory is
based on the following Heisenberg-like commutation relations between
coordinates:
\begin{equation} \label{cr}
[ \hat{x}^{\mu}, \hat{x}^{\nu}] =i \,\theta^{\mu \nu},
\end{equation}
where $\theta^{\mu\nu}$ is a constant antisymmetric matrix.

It is known that the construction of NC QFT in a general case
($\theta^{0i} \neq 0$) meets serious difficulties with unitarity and
causality \cite{GM} - \cite{CNT}. For this reason the version with
$\theta^{0i} = 0$ (space-space noncommutativity), in which there do
not appear such difficulties and which is a low-energy  limit of the
string theory, draws special attention. Then always there is a
system of coordinates, in which only $\thod = - \thdo \neq 0$. Thus,
when $\theta^{0i} = 0$, without loss of generality it is possible to
choose coordinates $\xz$ and $\xt$ as commutative and coordinates
$x^1$ and $x^2$ as noncommutative.

The relation (\ref{cr}) breaks the Lorentz invariance of the theory,
while the symmetry under the $SO \, (1,1) \times SO \, (2)$ subgroup
of the Lorentz group survives \cite{AB}. Translational invariance is
still valid.  Below we shall consider the theory to be $SO \, (1,1)$
invariant with respect to coordinates $\xz$ and $\xt$. Besides these
classical groups of symmetry, in the paper \cite{CKNT} it was shown,
that the noncommutative field theory with the commutation relation
(\ref{cr}) of the coordinates, and built according to the Weyl-Moyal
correspondence, has also a quantum symmetry, i.e. twisted
Poincar\'{e} invariance.

In the works \cite{AGM} - \cite{VM05} the Wightman approach was
formulated for NC QFT. For scalar fields the CPT theorem and the
spin-statistics theorem were proven in the case  $\theta^{0i} = 0$.

In \cite{AGM} it was proposed that Wightman functions in the
noncommutative case can be written down in the standard form
\begin{equation} \label{wf} \wf = \langle \, \Psi_0, \vf (x_1)
 \ldots \vf (x_n) \, \Psi_0 \, \rangle,
\end{equation}
where $\Psi_0$ is the vacuum state. However, unlike the commutative
case, these Wightman functions are only $SO \, (1,1) \otimes SO \,
(2)$ invariant.

In \cite{CMNTV} it was proposed that in the noncommutative case the
usual product of operators in the Wightman functions be replaced by
the Moyal-type product (see also \cite{Sz}):
\begin{equation} \label{mprod}
\vf (x_1) \star \cdots \star \vf (x_n) = \prod_{a<b} \,\exp{\left
({\frac{i}{2} \, \theta^{\mu\nu} \, \frac{ {\partial}}{\partial
x^{\mu}_a} \, \frac{ {\partial}}{\partial x^{\nu}_b}} \right)} \,\vf
(x_1)  \ldots  \vf (x_n), \; a, b = 1,2, \ldots n.
\end{equation}
Such a product of operators is compatible with the twisted
Poincar\'{e} invariance of the theory \cite{CPT} and also reflects
the natural physical assumption, that noncommutativity should change
the product of operators not only in coinciding points, but also  in
different ones. This follows also from another interpretation of NC
QFT in terms of a quantum shift operator \cite{CNT05}. In
\cite{VM05} it was shown that in the derivation of some axiomatic
results, the concrete type of product of operators in various points
is insignificant. It is essential only that from the appropriate
spectral condition (see formula (\ref{imp})), the analyticity of
Wightman functions with respect to the commutative variables $x^0$
and $x^3$ follows, while $x^1$ and $x^2$ do not need to be
complexified. The Wightman functions can be written down as follows
\cite{VM05}:
\begin{equation} \label{wfg}
 \wfts = \Pszl, \vfxtsp \Pszr.
\end{equation}
The meaning of $\bar\star$ depends on the considered case. In
particular,
$$
\varphi \, (x) \ts \varphi \, (y) = \varphi \, (x) \varphi \, (y) \;
 \mbox{according to \cite{AGM}} \qquad \mbox{or}
$$
$$
\varphi \, (x) \ts \varphi \, (y) = \varphi \, (x) \exp{\left
({\frac{i}{2} \, \theta^{\mu\nu} \, \frac{\overleftarrow
{\partial}}{\partial x^{\mu}} \, \frac{\overrightarrow
{\partial}}{\partial y^{\nu}}} \right)} \varphi \, (y) \; \mbox
{according to \cite{CMNTV}}.
$$
Note that actually field operators are the smoothed operators
\begin{equation} \label{vff}
\vff \equiv\int \,\vfx \,\fsx \, d \, x,
\end{equation}
where $\fsx$ are test functions. This point will be considered in
details in Section 2.

In \cite{VM05} it was shown that, besides the above-mentioned
theorems, in NC QFT (with $\theta^{0i} = 0$) a number of other
classical results of the axiomatic theory remain valid.  In
\cite{CPT} on the basis of the twisted Poincar\'{e} invariance of
the theory the Haag's theorem  was obtained \cite{HaagTh, FD} (see
also \cite{SW} and references therein).

In the present work, analogues of some known results of the
axiomatic approach in quantum field theory are obtained for the $S
\, O \, (1,1)$ invariant field theory, of which an important example
is NC QFT. In the $S \, O \, (1,3)$  invariant theory new
consequences of the generalized Haag's theorem are found, without
analogues in NC QFT.  At the same time it is proven that the basic
physical conclusion of Haag's theorem is valid also in the $S \, O
\, (1,1)$ invariant theory, and it is sufficient that spectrality,
local commutativity condition and translational invariance be
fulfilled only for the transformations concerning the commutating
coordinates. The analysis of Haag's theorem reveals essential
distinctions between commutative and noncommutative cases, more
precisely between the $S \, O \, (1,3)$ and $S \, O \, (1,1)$
invariant theories. In the commutative case, the conditions
(\ref{con1}) and (\ref{con2}), whose consequence is generalized
Haag's theorem, lead to the equality  of Wightman functions in two
theories up to four-point ones. In the present paper it is shown
that in the $S \, O \, (1, 1)$ invariant theory, unlike the
commutative case, only two-point Wightman functions are equal and it
is shown that from the equality of two-point Wightman functions in
two theories it follows that if in one of them the current is equal
to zero, it is equal to zero in the other as well and under weaker
conditions than the standard ones. It is also shown that for the
derivation of eq. (\ref{con2}) it is sufficient to assume that the
vacuum vector is a unique normalized vector, invariant under
translations along the axis $\xt$).  It is proven that from the
equality of four-point Wightman functions in two theories, the
equality of their elastic scattering amplitudes follows and, owing
to the optical theorem, the equality of total cross sections as
well. In derivation of this result LCC is not used. In the
noncommutative theory we also prove that classical results, such as
the irreducibility of the set of field operators, the theorems of
Reeh and Schlieder \cite{SW} - \cite{BLT} remain valid in the
noncommutative case. It should be emphasized that the results
obtained in this paper do not depend on the $S \, O \, (2)$
invariance of the theory in the variables $x^1$ and $x^2$ and
therefore can be extended to more general cases. The first theorem
of Reeh and Schlieder and the irreducibility of the set of field
operators remain valid in any theory, in which the spectral
condition (\ref{imp}) leads to the analyticity of the Wightman
functions in the variables $x^0$ and $x^3$ in the primitive domains
of analyticity ("tubes").

The study of Wightman functions leads still to new nontrivial
consequences also in the commutative case\footnote{Part of the
results have been presented in the talk \cite{Quarks}.}.

The paper is arranged as follows. In section 2 the necessary
properties of Wightman functions  are formulated; in section 3
generalizations of the theorems of Reeh and Schlieder to NC QFT are
obtained; in section 4 the irreducibility of the set of field
operators is proven; section 5 is devoted to generalized Haag's
theorem; in section 6 it is shown that in the commutative  case, the
conditions of weak local commutativity  (WLCC) and of local
commutativity (LCC), which are valid in the noncommutative case
((\ref{wftx}) and (\ref{wftxi})), appear to be equivalent to the
usual WLCC and LCC, respectively.

\section{Basic Properties of Wightman Functions }

In Wightman's approach cyclicity of the vacuum vector is assumed. In
the noncommutative case this means that any vector of the space
under consideration, $J$, can be approximated with arbitrary
accuracy by vectors of the type
\begin{equation} \label{vc}
\vffp  \, \Psi_0,
\end{equation}
where $f_i$ is a proper test function (\ref{vff}) and the
$\tilde\star$ in (\ref{vc}) acts only on the test functions
$f_1,...f_n$. For simplicity we consider the case of a real field,
however, the results are easily extended to a complex field. As well
as in the commutative case we assume
that the domain of definition of operators $\vff$ is dense in $J$\\
We admit that the scalar product of two vectors $\F = \vffkp  \,
\Psi_0$ and  $\Ps = \vffkpop$ is the following
$$
\langle \,\F, \Ps \,\rangle = \langle \, \Psi_0, \vffp  \,
\Psi_0\,\rangle =
$$
\begin{equation} \label{scprod}
 \int\, \wf \, \fsbxp \, \dxo \ldots
\dxn.
\end{equation}

For reasons given below, it is important only that the scalar
product of any two vectors in $J$ can be approximated by linear
combination of the scalar products of the type (\ref{scprod}) with
arbitrary precision. As noncommutativity does not affect the
commutative variables we assume that noncommutative Wightman
functions after smearing on noncommutative variables are tempered
distributions with respect to $x_i^0$ and $x_i^3$. As to their
properties with respect to noncommutative variables it is natural to
assume, as it is done in various versions of nonlocal theories
\cite{Luc} - \cite{Br}, that they belong to one of the
Gel'fand-Shilov spaces \cite{GSh}. This question will be considered
in a paper which is in preparation. Let us stress that the concrete
choice of Gel'fand-Shilov space is not important in the derivation
of our results.

Let us point out that the results of this work are new  also in the
case of the standard multiplication of functions $f_i\,(x_i)$.

For the results obtained below, translational invariance only in
commuting coordinates is essential, therefore we write down the
Wightman functions as:
\begin{equation} \label{ewf}
\wf = \wfksx,
\end{equation}
where $X$ designates the set of  noncommutative  variables $x_i^1,
x_i^2, \; i = 1, \ldots n$, and $\xi_j = \{\xi_j^0, \xi_j^3 \}$,
where $\xi_j^0 = x_j^0 - x_{j +1}^0, \xi_j^3 = x_j^3 - x_{j +1}^3$,
$j=1,...,n-1$.

Let us formulate now the spectral condition. We assume that any
vector in $p$ space, belonging to the complete system of these
vectors, is time-like with respect to momentum components  $\cpnz$
and $\cpnt$, i.e. that
\begin{equation} \label{imp}
\cpnz \geq | \cpnt |.
\end{equation}
The condition (\ref{imp}) is conveniently written as $\cpn \in
\bcvdpl$, where $\bcvdpl$ is the set of the four-dimensional vectors
satisfying the condition $P^0 \geq |P^3 |$. Recall that the usual
spectral condition for these vectors looks like $\cpn \in \bcvpl$,
i.e. $\cpnz \geq | \vcpn | $. From the condition (\ref{imp}) and the
completeness of the system of the vectors $\Psi_{\cpn}$:
\begin{equation} \label{last}
\langle\Phi, \Psi\rangle = \sum_n\int d \,\cpn \langle\Phi, \Psi
_{\cpn} \rangle \langle\Psi_{\cpn}, \Psi\rangle,
\end{equation}
it follows that
\begin{equation} \label{10}
\int \, d \, a \, e^{- i \, p \, a} \, \langle\Phi, U \, (a) \,
\Psi\rangle = 0, \quad \mbox{if} \quad p \not \in \bar V_2^+,
\end{equation}
where $a = \{a^0, a^3 \}$ is a two-dimensional vector, $U \, (a)$ is
a translation in the plane $p^0, p^3$, and $\F$ and $\Ps$ are
arbitrary vectors. The equality (\ref{10}) is similar to the
corresponding equality in the standard case (\cite{SW}, Chap.~2.6).
Let us point out that in fact the complete system of vectors
$\Psi_{\cpn}$ contains vectors with infinite norms. But this point
can be settled in the noncommutative case just as in commutative one
\cite{SW}. The point is that noncommutativity does not affect the
momentum operators. So in momentum space the only difference between
commutative and noncommutative cases is the weaker spectral
condition in the latter case.

A direct consequence of the equality (\ref{10}) is the spectral
property of Wightman functions:
\begin{equation} \label{wfsp}
W\, (P_1..., P_{n - 1}, X) = \frac{1}{(2\pi)^{{n - 1}}} \, \int \,
e^{i \, P_j \,\xi_j} \, W \, (\xi_1..., \xi_{n - 1}, X) \, d
\,\xi_1 ... d \,\xi_{n - 1} = 0,
\end{equation}
if $P_j \not \in \bar V_2^+$. The proof of the equality (
\ref{wfsp}) is similar to the proof of the spectral condition in the
commutative case \cite{SW}, \cite{BLT}. Recall that in the latter
case the equality (\ref{wfsp}) is valid, if $P_j \not \in \bar V^+$.
Having written down $\wfksx$ as
\begin{equation} \label{wfspo}
W\, (\xi_1..., \xi_{n - 1}, X) = \frac{1}{(2\pi)^{{n - 1}}} \,
\int \, e^{- i \, P_j \,\xi_j} \, W \, (P_1..., P_{n - 1}, X) \, d
\, P_1 ... d \, P_{n - 1},
\end{equation}
and taking into  account that Wightman functions are tempered
distributions with respect to the commutative variables, we obtain
that, due to the condition (\ref{wfsp}), $\wfnx$ is analytical in
the "tube" $\; \ctnmi$:
\begin{equation} \label{tube}
\nui \in \ctnmi, \quad \mbox{if} \quad \nui = \ksi - i \,\eti, \;
\eti \in V_2^+, \; \eti = \{\etiz, \etit \}.
\end{equation}
It should be stressed that the noncommutative coordinates $\xito, \;
\xitd$ remain always real.

Owing to  $S \, O \, (1,1)$ invariance and according to the
Bargmann-Hall-Wightman theorem \cite{SW} - \cite{BLT}, $\wfnx$ is
analytical in the domain $\ctn$
\begin{equation} \label{ctn}
\ctn = \cup_{\Lac} \, \Lac \,\ctnmi,
\end{equation}
where $\Lac \in S \, O_c \, (1, 1)$ is the two-dimensional analogue
of the complex Lorentz group. This expansion is similar to the
transition from tubes to expanded tubes in the commutative case.
Just as in the commutative case, the expanded domain of analyticity
contains real points $\xit$, which are the noncommutative Jost
points, satisfying the condition $\xit \sim \xj, \; \forall \; i,
j$, which means that
\begin{equation} \label{app}
{ \left (\xitz - \xjz\right)}^2 - {\left (\xitt - \xjt\right)}^2 <
0.
\end{equation}
It should be emphasized that the noncommutative Jost points are a
subset of the set of Jost points of the commutative case, when
\begin{equation} \label{pp}
{ \left (\xit - \xj\right)}^2 < 0 \qquad \forall \; i, j.
\end{equation}
Let us point out that WLCC and LCC, respectively have the same form
as in the local theory:
\begin{equation} \label{wftx}
\wf = \wfo, \quad \mbox{if} \quad \xit \sim \xj  \quad \forall \; i,
j;
\end{equation}
\begin{equation} \label{wftxi}
\wfxi = \wfxio,
\end{equation}
if supp $\fsi \in O_i \times R^2$, supp $\fsipo \in  O_{i+ 1} \times
R^2, \;   O_i \sim O_{i+ 1}$, which means that the condition
(\ref{app}) is valid for any points $x_i \in O_i \times R^2$ and
$x_{i+ 1} \in O_{i+ 1} \times R^2$.

\section{Theorems of Reeh and Schlieder in NC QFT}

In the following we shall prove the analogues of the theorems of
Reeh and Schlieder \cite{SW, Jost} for the noncommutative case.

{\bf  Theorem 1} \quad{\it Let supports of functions $\fsi$ belong
to $O \times R^2$, where $O$ is any open domain on  variables $\xiz$ and $\xitr$. \\
Then there is no vector distinct from zero, which is orthogonal to
all vectors of the type $\vffp  \, \Psi_0$, supp $\fsi \in O \times
R^2$.}
Let us consider two vectors
\begin{eqnarray} \label{vec}
\Phi & = & \vffp  \, \Psz, \quad \mbox{supp} \quad \fsi \in O \times R^2 \quad \forall \; i,  \nonumber \\
\Psi & = &  \vftfm\,\ts \,\ldots \ts \, \vftfo\, \Psz.
\end{eqnarray}
On $supp\ \tilde {f_i}$ no restrictions are imposed. We shall prove
that $\Psi = 0$, if for any vector $\Phi$
\begin{equation} \label{17}
\Psl, \Phr = 0.
\end{equation}
For the proof it is sufficient to notice that the corresponding
Wightman function
$$
\Pszl, \vfytsp \,\ts \,\vfxtsp \,\Pszr \equiv
W_{\tilde\star}(y_1,...y_m,x_1,...,x_n)
$$
is an analytical function in the variables $- x_1^0 - i\,\eta_0^0,
\; - x_1^3 - i \,\eta_0^3, \; \nui = \ksi - i \,\eti, \; i = 1,
\ldots n - 1$, if $\eti \in \cvdpl$. According to the condition
(\ref{17}), this function is equal to zero on the border, if $\xit
\in O \times R^2$. As $O$ is an open domain,
$W_{\tilde\star}(y_1,...y_m,x_1,...,x_n) \equiv 0$. Thus the vector
$\Psi$ is orthogonal to all vectors of the type (\ref{vc}) and,
according to the cyclicity of the vacuum vector, $\Psi = 0$. Taking
into account the completeness of the system of vectors $\Psi$ we
come to the statement of the Theorem. Remark that for the proof of
the Theorem 1 only the analyticity of the Wightman functions in the
domain $\ctnmi$ has been used.

{ \bf Theorem 2} \quad {\it Let the support of $\tilde f \in \tilde
O \times R^2$, where $\tilde O$ is such a domain of commutative
variables, for which domain $O \sim \tilde O$, satisfying the
condition of the Theorem 1, exists. Then the condition
\begin{equation} \label{18}
\vftf  \, \Psz = 0
\end{equation}
implies that
\begin{equation} \label{19}
\vftf  \equiv 0,
\end{equation}
if the operator $\vftf$ satisfies the LCC.  }

In accordance with LCC
\begin{equation} \label{21}
\vftf \, \ts \,\Phi = 0,
\end{equation}
if vector $\Phi$ is defined as in eq. (\ref{vec}). Hence, for any
vector $\Psi$ belonging to the domain of definition of the Hermitian
operator $\vftf$,
\begin{equation} \label{22}
\langle \vftf  \, \ts  \, \Psi, \Phi \rangle = \langle \, \Psi,
\vftf  \, \ts \, \Phi \rangle = 0.
\end{equation}
According to the Theorem 1, the condition (\ref{22}) means that
$\vftf \, \ts \, \,\Psi = 0$. As the domain of definition of the
operator $\vftf$ is dense in $J$, this equality means the validity
of the equality (\ref{19}).

{ \bf  Remark} \quad{\it Theorem 2 remains true for any densely
defined operator $\pstf$, mutually local with $\vff$, i.e. if
\begin{equation} \label{23}
\pstf \ts \vff \ts \Phi = \vff \ts \pstf \ts \Phi,
\end{equation}
if $supp \; f \in O \times R^2, \; supp\ \tilde f \in \tilde O
\times R^2, \;  O \sim \tilde O$, vector  $\Phi$ belongs to the
domain of definition of operators $\vff$ and $\pstf$. }

\section{Irreducibility of the set of field operators $\vff$ in NC QFT}

In  the noncommutative case, the irreducibility of a set of field
operators $\vff$ implies that, from the condition
\begin{equation} \label{24}
A \, \vffp  \, \Psi_0 =  \vffp  \, A \,  \Psi_0
\end{equation}
where $\fsi$ are arbitrary test functions and $A$ is a
bounded operator, follows that
\begin{equation} \label{25}
A = C \,\identity \, \qquad C \in \complessi \,
\end{equation}
where $\identity$ is the identity operator.

In the noncommutative case the condition of irreducibility of the
set of operators $\vff$ is  valid as well as in commutative case.
The point is that for this it is sufficient to have the
translational invariance in the variable $\xz$ and the spectral
condition, which can be weakened up to the condition
\begin{equation} \label{26}
\cpnz \geq 0.
\end{equation}
Using condition (\ref{24}) and the invariance of the vacuum vector
with respect to the translations $U \, (a)$ on the axis $\xz$, we
obtain the following equality
\begin{equation} \label{27}
\langle \, A^* \,\Psz, U \, (a) \, \vffp \,\Psz \,\rangle =
\langle \, \vffnp \,\Psz, U \, (-a) \, A \,\Psz \,\rangle.
\end{equation}
In accordance with the eq. (\ref{10})
$$
\int \, d \, a \, e ^{- i \, p^0 \, a} \, \langle \, A ^* \,\Psz,
U\, (a) \, \vffp  \,\Psz \,\rangle \neq 0,
$$
only if $p^0 \geq 0$. However,
$$
\int \, d \, a \, e ^{- i \, p^0 \, a} \, \langle \,\vffp  \,\Psz, U
\, (- a) \, A \,\Psz \,\rangle \neq 0,
$$
only if $p^0 \leq 0$. Hence, the equality (\ref{27}) can be
fulfilled only when $ p^0 = 0$ and so, in accordance with the eq.
(\ref{last}),
\begin{equation} \label{28}
A\,\Psz = C \,\Psz.
\end{equation}
Thus owing to (\ref{24}) and (\ref{28})
\begin{equation} \label{29}
A\,\vffp  \,\Psz = C \,\vffp  \,\Psz.
\end{equation}
The required equality (\ref{25}) follows from eq. (\ref{29}) in
accordance with the boundedness  of the operator $A$ and ciclicity
of the vacuum vector.

\section{Generalized Haag's Theorem }

Recall the formulation of the generalized Haag's theorem in the
commutative case (\cite{SW}, Theorem~4.17):

{\it Let $\vf_f^1 \, (t)$ and $\vf_f^2 \, (t), \, supp\ f \in R^3$
be two irreducible sets of operators, for which the vacuum  vectors
$\Psi_0^1$ and $\Psi_0^2$ are cyclic. Further, let the corresponding
Wightman functions be analytical in the domain
$\ctn$\footnote{Remark that the required analyticity of the Wightman
functions
follows only from the spectral condition and the $S \, O (1, 3)$ invariance of the theory.}. \\
Then the two-, three- and four-point Wightman functions coincide in
the two theories if there is a unitary operator $V$, such that}
\begin{eqnarray}
&&1)\ \ \vf_f^2 \, (t) = V \,\vf_f^1 \, (t) \, V ^ *,
\label{con1}\\
&&2)\ \ \Psi_0^2 = C \, V \,\Psi_0^1, \quad C \in \complessi, \quad
|C | = 1.\label{con2}
\end{eqnarray}

It should be emphasized that actually the condition  $2)$ is a
consequence of condition $1)$ with rather general assumptions (see
the {\bf Statement} below). In the formulation of  Haag's theorem it
is assumed that the formal operators $\vfitvx$ can be smeared only
on the spatial variables. This assumption is natural also in
noncommutative case if $\theta^{0i} =0 $.

Let us consider Haag's theorem in the $S \, O (1, 1)$  invariant
field theory and show that the corresponding equality is true only
for two-point Wightman functions.

For the proof we first note that in the noncommutative case, just as
in the commutative one, from conditions  $1)$ and  $2)$ it follows
that the Wightman functions in the two theories coincide at equal
times
\begin{equation} \label{32}
\langle \Psi_0^1, \vf_1 \, (t, \vec{x_1}) \, \ts \cdots \vf_1 \,
(t, \vec{x_n}) \, \Psi_0^1 \rangle = \langle \Psi_0^2, \vf_2 \,
(t, \vec{x_1}) \, \ts \cdots \vf_2 \, (t, \vec{x_n}) \, \Psi_0^2
\rangle.
\end{equation}

Having written down the two-point Wightman functions $W_i \, (x_1,
x_2), \; i = 1, 2$ as  $W_i \, (u_1, v_1; u_2, v_2)$, where $u_i =
\{x_i^0, x_i^3 \}, \; v_i = \{x_i^1, x_i^2 \}$ we can write for them
equality (\ref {32}) as:
\begin{equation} \label{32a}
W_1 \, (0, \xi^3; v_1, v_2) = W_2 \, (0, \xi^3; v_1, v_2),
\end{equation}
where $\xi = u_1 - u_2, \; v_1$ and $v_2$ are arbitrary vectors. Now
we  notice that, due to the $S \, O (1, 1)$ invariance,
\begin{equation} \label{32b}
W_i \, (0,{\xi} ^3; v_1, v_2) = W_i \, (\tks; v_1, v_2)
\end{equation}
hence,
\begin{equation} \label{32c}
W_1 \, (\tks; v_1, v_2) = W_2 \, (\tks; v_1, v_2),
\end{equation}
where $\tks$ is any Jost point. Due to the analyticity of the
Wightman functions in the commuting variables they are completely
determined by their values at the Jost points. Thus at any $\xi$
from the equality (\ref{32c}), it follows that
\begin{equation} \label{32d}
W_1 \, (\xi; v_1, v_2) = W_2 \, (\xi; v_1, v_2).
\end{equation}
As $v_1$ and $v_2$ are arbitrary, the formula (\ref{32d}) means the
equality of two-point Wightman functions at all values of arguments.

Thus, for the equality of the two-point Wightman functions in two
theories related by the conditions (\ref{con1}) and (\ref{con2}),
the $S \, O (1, 1)$ invariance of the theory and corresponding
spectral condition are sufficient.

It is impossible to extend this proof to three-point Wightman
functions. Indeed, let us write down $W_i \, (x_1, x_2, x_3)$ as
$W_i \, (u_1, u_2, u_3; v_1, v_2, v_3)$, where vectors $u_i$ and
$v_i$ are determined  as before. Equality (\ref{32a}) means that
\begin{equation} \label{32e}
W_1 \, (0, \xi_1^3, 0, \xi_2^3; v_1, v_2, v_3) = W_2 \, (0, \xi_1^3,
0, \xi_2^3; v_1, v_2, v_3),
\end{equation}
$v_1, v_2, v_3$ are arbitrary. In order to have equality of the
three-point Wightman functions in the two theories from the $S \, O
(1, 1)$ invariance, the existence of transformations $\Lambda \in S
\, O (1, 1)$ connecting the points $(0, \xi_1^3)$ and $(0, \xi_2^3)$
with an open vicinity of Jost points is necessary. That would be
possible, if there existed two-dimensional vectors $\tilde{\xi}_1$
and $\tilde{\xi}_2$, $(\tilde{\xi}_i = \Lambda \, (0, \xi_i^3))$,
satisfying the inequalities:
$$
{(\tilde{\xi}_1)} ^2 < 0, \quad{(\tilde{\xi}_2)} ^2 < 0, \quad |
(\tilde{\xi}_1, \tilde{\xi}_2) | < \sqrt{{(\tilde{\xi}_1)}^2
{(\tilde{\xi}_2)}^2}.
$$
These inequalities are similar to the corresponding inequalities in
the commutative case (see equation (4.87) in \cite{SW}). However, it
is easy to check that the last of these inequalities can not be
fulfilled, while the first two are fulfilled.

Let us show now that the condition (\ref{con2}) actually is a
consequence of the condition (\ref{con1}).

{\bf  Statement} \quad{\it  Condition (\ref{con2}) is fulfilled, if
the vacuum vectors $\Pszi$ are unique, normalized, translationally
invariant vectors with respect to translations $U_i \, (a)$ along
the axis $\xt$.}

It is easy to see that the operator $U_1^{- 1} \, (a) \, V ^{- 1} \,
U_2 \, (a) \, V$ commutes with operators $\vf_f^1 \,(t)$ and, owing
to the irreducibility of the set of these operators, it is
proportional to the identity operator. Having considered the limit
$a = 0$, we see that
\begin{equation} \label{34}
U_1 ^{ - 1} \, (a) \, V ^{- 1} \, U_2 \, (a) \, V = \identity.
\end{equation}
From the equality (\ref{34}) it follows directly that if
\begin{equation} \label{35}
U_1 \, (a) \, \Pszo = \Pszo,
\end{equation}
then
\begin{equation} \label{36}
U_2 \, (a) \, V \,\Pszo = V \,\Pszo,
\end{equation}
i.e. the condition (\ref{con2}) is fulfilled. If the theory is
translationally invariant in all variables, the equality (\ref{36})
is true, if the vacuum vector is unique, normalized, translationally
invariant in the spatial coordinates.

The most important consequence of the generalized  Haag theorem is
the following statement: if one of the two fields related by
conditions (\ref{con1}) and (\ref{con2}) is a free field, the other
is also free. In  deriving this result the equality of the two-point
Wightman functions in the two theories  and LCC are used. In
\cite{CPT} it is proved that this result is valid also in the
noncommutative theory, if $\theta^{0i} = 0$.

Here we obtain the close result in the $S \, O (1, 1)$ symmetric
theory using the spectral conditions  and translational invariance
only with respect to the commutating coordinates.  In this case the
equality of the two-point Wightman functions in the two theories
leads to the conclusion that if LCC (\ref{wftxi}) is fulfilled and
the current in one of the theories is equal to zero, for example,
$j_f^1 = 0$, then $j_f^2 = 0$ as well; $\; j_f^{i} = (\square +
m^{2}) \, \varphi_f^{i}$. Indeed as $W_{1} \, (\xo, \xd) = W_{2} \,
(\xo, \xd)$,
\begin{equation} \label{43}
< \Psi_0^1,  j^1 \, (\xo) j^1 \, (\xd) \, \Psi_0^1 > = < \Psi_0^2,
j^2 \, (\xo) j^2 \, (\xd) \, \Psi_0^2> = 0,
\end{equation}
since $j_f^1  = 0$. Hence,
$$
j_f^2  \, \Psi_0^2 = 0.
$$
Here we assume that  $J$ is a positive metric space.  It is
sufficient to take advantage of the Theorem 2 from which follows
that $j_f^2 = 0$ (see the Remark after {\bf Theorem 2}), since LC
implies mutual local commutativity of a field operator and the
corresponding current.

Let us proceed now to the $S \, O (1, 3)$ symmetric theory. In this
case we show  that from the equality of the four-point Wightman
functions for the fields ${\varphi}_f^1 \, (t)$ and ${\varphi}_f^2
\, (t)$, related by the conditions (\ref{con1}) and (\ref{con2}),
which takes place in the commutative theory, an essential physical
consequence follows. Namely, for such fields the elastic scattering
amplitudes of the corresponding theories coincide, hence, due to the
optical theorem, the total cross-sections coincide as well. In
particular, if one of these fields, for example, ${\varphi}_f^1$ is
a trivial field, i.e. the corresponding $S$ matrix is equal to
unity, also the field ${\varphi}_f^2$ is free. In the derivation of
this result the local commutativity condition is not used. The
statement follows directly from the Lehmann-Symanzik-Zimmermann
reduction formulas \cite{LSZ}. Here and below dealing with the
commutative case in order not to complicate formulas we consider
operators ${\varphi}_1 \, (x)$ and ${\varphi}_2 \, (x)$ as they are
given in a point.

Let $< p_3, p_4 | p_1, p_2 >_{i}, \; i = 1,2$ be an elastic
scattering amplitudes  for the fields ${\varphi}_1 \, (x)$ and $
{\varphi}_2 \, (x)$ respectively. Owing to the reduction formulas,
$$
< p_3, p_4 | p_1, p_2 >_{i} \, \sim \, \int \,\dxo \cdots \dxch \, e
^{i \, (-p_1 \,\xo - p_2 \,\xd + p_3 \,x_3 + p_4 \,\xch)} \, \cdot
$$
\begin{equation} \label{41}
\prod_{j = 1} ^{4} \, (\square_{j} + m ^{2}) \, < 0 |
T\,\varphi_{i} \,
(\xo) \, \cdots \, \varphi_{i} \, (\xch) | 0 >,
\end{equation}
where $T \,\varphi_{i} \, (\xo) \, \cdots \, \varphi_{i} \, (\xch)$
is the chronological product of operators. From the equality
$$
W_2 \, (\xo, \ldots, \xch) = W_1 \, (\xo, \ldots, \xch)
$$
it follows that
\begin{equation} \label{42}
< p_3, p_4 | p_1, p_2 >_{2} = < p_3, p_4 | p_1, p_2 >_{1}
\end{equation}
for any $p_{i}$. Having applied this equality for the forward
elastic scattering amplitudes, we obtain that, according to the
optical theorem, the total cross-sections for the fields
${\varphi}_1 \, (x)$ and ${\varphi}_2 \, (x)$ coincide. If now the
$S$-matrix for the field ${\varphi}_1 \, (x)$ is unity, then it is
also unity for field ${\varphi}_2 \, (x)$. We stress that  the
equality of the four-point Wightman functions in the two theories
related by the conditions (\ref{con1}) and (\ref{con2}) are valid
only in the commutative field theory but not in the noncommutative
case.

\section{Equivalence of various
 conditions of local commutativity in QFT}

Let us show that in the commutative case, when Wightman functions
are analytical ones in the usual domain, the conditions (\ref{wftx})
and (\ref{wftxi}) are equivalent to the standard conditions of WLC
and LC, i.e. the latter remain valid if the condition (\ref{pp}) is
fulfilled. In effect, (\ref{wftx}) is  a sufficient condition for
the theory to be CPT invariant \cite{AGM}. However, in the
commutative case, from CPT invariance the standard condition of WLC
follows. \cite{SW} - \cite{BLT}.

The equivalence of LCC (\ref{wftxi}) with the standard one follows
from the fact that, for the validity of usual LCC its validity on
arbitrary small spatially divided domains is sufficient (see \cite
{BLOT}, Proposal 9.12). Indeed, validity of "noncommutative" LCC
(\ref{wftxi}) in commutative the case means validity of standard LCC
in the domain ${(x^{0} - y^{0})}^2 - {(x^{3} - y^{3})}^2 < 0, \;
x^{k}, y^{k}, \; k = 1, 2$ are arbitrary. This domain satisfies the
requirements of the above mentioned statement.

Besides we can replace (\ref{wftxi}) with the formally weaker
condition, requiring that it is valid only when
\begin{equation} \label{lsq}
{\left (\xitz - \xjz \right)}^2 - {\left (\xitt - \xjt\right)}^2 < -
l^{2}, \; \forall \; i, j,
\end{equation}
where $l$  is any fixed fundamental length. Indeed, in the
commutative theory, according to the results of Wightman, Petrina
and Vladimirov (see \cite{Vlad}, Chapter~5 and references therein)
the condition
\begin{equation} \label{45}
[\varphi \, (x), \varphi \, (y)] = 0, \quad (x - y)^{2} < - l^{2},
\end{equation}
for any finite $l$, is equivalent to standard LCC $(l = 0)$.
Similarly if (\ref{wftxi}) is fulfilled at (\ref{lsq}), then it is
fulfilled also  at $l = 0$.

Thus, the analysis of Wightman functions in NC QFT, carried out in
this and our previous works \cite{CMNTV}, \cite{VM05}, \cite{CPT},
shows that the basic axiomatic results are valid (or have analogues)
in NC QFT as well, at least in the case when $\theta^{0i} = 0$.


\begin{thebibliography}{99}

          \bibitem{SW}
 R. F. Streater and A. S. Wightman, {\it PCT, Spin and Statistics and
All That}, Benjamin, NewYork (1964).

          \bibitem{Jost}

2. R. Jost, {\it The General Theory of Quantum Fields },Amer.
Math.Soc., Providence, R.I. (1965).

          \bibitem{BLT}
 N. N. Bogoliubov, A. A. Logunov, and I. T. Todorov,
{\it Introduction to Axiomatic Quantum Field Theory}, Benjamin,
Reading, Mass (1975).

          \bibitem{BLOT}
 N. N. Bogoliubov, A. A. Logunov, A. I. Oksak and I. T. Todorov,
{\it General Principles  of Quantum Field Theory}, Kluwer, Dordrecht
(1990).

          \bibitem{Haag}
R. Haag, {\it Local Quantum Physics}, Springer, Berlin (1996).


          \bibitem{DN}
M. R. Douglas and N. A. Nekrasov, {\it Rev. Mod. Phys.} {\bf 73} 977
(2001), hep-th/0106048.

          \bibitem{Sz}
R. J. Szabo,{\it Phys. Rept.} {\bf 378} 207 (2003), hep-th/0109162.

          \bibitem{Snyder}
H. S. Snyder, {\it Phys. Rev.} {\bf 71} 38 (1947).

          \bibitem{Connes}
A. Connes, {\it Noncommutative Geometry},  Academic Press, New
York (1994).

          \bibitem{DFR}
S. Doplicher, K. Fredenhagen and J. E. Roberts, {\it Phys. Lett. B}
{\bf 331} 39 (1994); {\it Comm. Math. Phys.}, {\bf 172}, 187 (1995).

          \bibitem{SeWi}
N. Seiberg and E. Witten, {\it JHEP} {\bf 9909} 32 (1999),
hep-th/9908142.

         \bibitem{ABZ}
L. \'Alvarez-Gaum\'e, J. L. F. Barbon and R. Zwicky, {\it JHEP} {\bf
0105} 057 (2001), hep-th/0103069.

           \bibitem{GM}
J. Gomis and T. Mehen, {\it Nucl. Phys. B} {\bf 591} 265 (2000),
hep-th/0005129.

          \bibitem{SSB}
N. Seiberg, L. Susskind and N. Toumbas, {\it JHEP} {\bf 0006} 044
(2000), hep-th/0005015.

         \bibitem{AB}
L. \'Alvarez-Gaum\'{e} and J. L. F. Barbon, {\it Int. J. Mod. Phys.
A}, {\bf 16} 1123 (2001), hep-th/0006209.

           \bibitem{CNT}
M. Chaichian, K. Nishijima and A. Tureanu, {\it Phys. Lett. B} {\bf
568} 146 (2003), hep-th/0209006.


           \bibitem{CKNT}
M. Chaichian, P.P. Kulish, K. Nishijima and A. Tureanu, {\it Phys.
Lett.} {\bf B 604} 98, hep-th/0408069.

          \bibitem{AGM}
L. \'Alvarez-Gaum\'{e} and M. A. V\'azquez-Mozo, {\it Nucl. Phys.
B}, {\bf 668} 293 (2003), hep-th/0305093.

          \bibitem{CMNTV}
M. Chaichian, M. N. Mnatsakanova, K. Nishijima, A. Tureanu and Yu.
S. Vernov, hep-th/0402212.

         \bibitem{VM05}
Yu.S. Vernov, M.N. Mnatsakanova, {\it  Theor. Math. Phys.} {\bf 142}
337 (2005).

          \bibitem{CPT}
M. Chaichian, P. Pre\v snajder and A. Tureanu, {\it Phys. Rev.
Lett.}, {\bf 94} 151602 (2005), hep-th/0409096.


          \bibitem{CNT05}
M. Chaichian, K. Nishijima and A. Tureanu, {\it Phys. Lett.}, {\bf B
633} 129 (2006), hep-th/0511094.


          \bibitem{HaagTh}
R.Haag, {\it Dan. Mat. Fys.} {\bf 29} 12 (1955).

          \bibitem{FD}
P.G. Federbush and K.A. Jonson, {\it Phys. Rev.} {\bf 120} 1926
(1960); R. Jost, {\it Properties of Wightman Functions}, edited by
E.R. Caianiello, Lectures on Field Theory and Many-Body Problem,
Academic Press, New Jork, (1961).

\bibitem{Quarks}
M. Chaichian, M. Mnatsakanova, A. Tureanu and Yu. Vernov, Talk given
at the 14th International Seminar on High Energy Physics
QUARKS'2006, Repino, St.Petersburg, Russia, May 19-25, 2006,
hep-th/0611097.

          \bibitem{Luc}
W. L\"{u}cke, {\it J. Math. Phys} {\bf 27} 1901 (1985).

          \bibitem{Sol}
M. A. Solov'ev, {\it  Theor. Math.Phys. } {\bf 121} 1377 (1999).

          \bibitem{Br}
E. Br\"{u}ning and S. Nagamachi, {\it  J. Math. Phys.} {\bf 30} 2340
(1989).

          \bibitem{GSh}
Gel'fand and Shilov, {\it Generalized Functions}, V.2, Chapter IV,
Academic Press Inc., New York (1968).

          \bibitem{LSZ}

H. Lehmann, K. Symanzik, W. Zimmermann,  {\it  Nuovo Cim.} {\bf 1}
205 (1955); {\it  Nuovo Cim.}, {\bf 6}, 319 (1957).

          \bibitem{Vlad}
A.S. Wightman,   {\it J. Indian Math. Soc.}, {\bf 24} 625
(1960-61);\\
D. Ya. Petrina,  {\it Ukr. Mat. Zh.} {\bf 13}, 109 (1961);\\
V.S. Vladimirov, {\it Sov. Math. Dokl.} {\bf 1} (1960) 1039; {\it
Methods of the Theory of Functions of Several Complex Variables},
Cambridge, Massachusetts, MIT Press, 1966.

\end{thebibliography}
\end{document}